# Are we ready for spray-on carbon nanotubes?

As artists and manufacturers explore the use of spray-on carbon nanotube coatings, Andrew Maynard explores the state of the science around nanotube safety.

**Andrew D. Maynard**

*Director, Risk Innovation Lab. School for the Future of Innovation in Society, Arizona State University, Tempe AZ, USA. Email: Andrew.maynard@asu.edu*

Earlier this year, British sculptor, Anish Kapoor was given exclusive rights to use a new spray-on carbon nanotube-based paint. The material – produced by UK-based Surrey NanoSystems and marketed as Vantablack S-VIS – can be applied to a range of surfaces, and absorbs well over 99% of the light that falls onto it.[1] It is claimed to be the world's blackest paint, and there is growing interest in its use in works of art and high-end consumer products.

It's easy to see the appeal of Vantablack S-VIS.  Apart from technical applications where stray reflections need to be suppressed, this is a material that potentially enables manufacturers and artists to give their products a unique aesthetic edge.  Yet, having worked on carbon nanotube safety for some years, I was intrigued to see the material in a spray-paint designed to coat objects that people may possibly come into contact with. It was, after all, only a few years ago that journalists were asking if carbon nanotubes were the next asbestos.[2]  And while this is unlikely, concerns over the possible health impacts of the material persist.

Safety concerns of purposely manufactured carbon nanotubes began in the 1990's. Within 12 months of the in-depth description of this material by Sumio Iijima and colleagues[3], Gerald Coles (an occupational hygiene expert) raised concerns over the potential asbestos-like health impacts of this fibrous material.[4]  At the time, his concerns did not gain much attention.  Yet, a little over a decade later, carbon nanotube safety was increasingly being scrutinized.

Ten years after Coles' letter, I was part of a collaboration studying the potential risks of single walled carbon nanotubes that were being synthesized in two research laboratories.  Our research showed both the potential for exposure to clumps of nanotubes while handling the material,[5] and their potential to lead to granulomas and interstitial fibrosis in the lungs.[6]

While these and other early safety studies were far from conclusive, they did raise concerns over unanticipated and possibly serious health and environmental risks from carbon nanotubes. These were reflected in a seminal 2004 review of nanotechnology risks and benefits by the UK Royal Society and Royal Academy of Engineering, which recommended that "the release of manufactured nanoparticles and nanotubes into the environment be avoided as far as possible" until more is known about their environmental impacts.[7]

As research progressed, it quickly became clear that the number of graphene-like layers, or "walls" that make up these nanotubes, had a profound influence on both their physicochemical properties and their potential biological impacts. As a result, from around 2005, two distinct research paths began to emerge: one examining the health risks of convoluted and tangled single-walled carbon nanotubes, and another looking at more fiber-like multi-walled carbon nanotubes.

Multi-walled carbon nanotubes were of particular concern, as – echoing Coles' comments in 1992 – they appeared to share a number of properties with harmful asbestos fibers, including (in some cases) being long, thin, and durable. Reflecting this, fourteen researchers (myself included) recommended in 2006 that "the potential health impact of high-aspect-ratio, biopersistent engineered nanotubes, nanowires and nanofibres is systematically investigated within the next 5 years."[8]

Spurred on by this recommendation, Ken Donaldson and colleagues published a landmark study in 2008 exploring similarities in health risk between carbon nanotubes and asbestos.[9] Using peritoneal injection in mice, they demonstrated the potential for long, straight (but importantly, not short, curled and/or tangled) multi-walled carbon nanotubes to initiate a chain of biological events that could lead to mesothelioma - a cancer of the tissue surrounding the lungs with poor prognosis that is predominantly associated with asbestos exposure. Donaldson's work was paralleled by research in Japan led by Jun Kanno, that further supported the hypothesis that long, thin, straight multi-walled carbon nanotubes may behave like asbestos.[10] Since then, a growing body of research has indicated that inhaled carbon nanotubes that are long (more than ~10 – 20 μm in length), straight, durable, and not part of a larger aggregate, have the potential to cause cancers of the lung and surrounding tissues, including mesothelioma.[11]

However, not all carbon nanotubes look and act like asbestos fibers. They can vary in their number of graphene-like walls; their length, width, and curvature; their crystallinity; the impurities (such as catalyst nanoparticles) attached to them; substances that intentionally coat them or are embedded in them; and the degree to which they are locked into a larger matrix of nanotubes and other materials.  Each combination of physical and chemical characteristics – and there are tens of thousands of such combinations – alters the potential of carbon nanotubes to get into and move through the human body, and to cause harm.

Understanding how different combinations of characteristics and exposure scenarios affect potential health risks has proved to be a herculean task, and it is still hard to predict how any given sample of carbon nanotubes will impact someone if they are exposed.  But researchers are slowly piecing together the puzzle that represents carbon nanotube toxicity and risk.

In early 2000, the Japanese trading company Mitsui & Co. Ltd. decided to develop a commercial scale multi-walled carbon nanotube plant.  The technology they purchased was based on a continuous production process that led to particularly pure and well-defined multi-walled carbon nanotubes that were thin, straight, and ranged in length from a few micrometers to around 20 micrometers.  Concerned about the safety of their product, Mitsui began exploring the toxicity of their nanotubes. And in 2005, they freely provided samples to toxicologists in a number of research labs – including those at the US National Institute for Occupational Safety and Health (NIOSH).

As NIOSH researchers got close to publishing their first studies on the Mitsui nanotubes, they realized they needed a way to refer to this material.  In the absence of further guidance, they took the name from the FedEx invoice that came with the shipment: "MWNT-7".  The "7" as it turned out, was an arbitrary shipping identification number assigned by a company employee – chosen because "he loved Lucky-7 for gambles".[12]  The name stuck, and MWNT-7 (and sometimes MWCNT-7 or Mitsui-7) became one of the most widely studied – if, ironically, not the luckiest – forms of multi-walled carbon nanotubes.

In 2014, the International Agency for Research on Cancer (IARC) published an evaluation of the carcinogenic potential of carbon nanotubes.  Unfortunately, because of the variety of carbon nanotube types used in published toxicity studies, the review panel could not draw generalizable conclusions that satisfied IARC's rigorous standards.  The panel concluded that carbon nanotubes "cannot be classified due to a lack of data" (group 3 carcinogen). All, that is, apart from MWNT-7. Because MWNT-7 had been studied so extensively, there was sufficient evidence for the panel to place this one specific form of carbon nanotubes into group 2B – "possibly carcinogenic to humans".[13]

The following year, Günter Oberdörster and co-authors published what is possibly the most comprehensive review of carbon nanotube toxicology studies to date.[14] Focusing on inhalation of nanotubes, they document evidence of transient pulmonary inflammation and, rapid and persistent development of granulomatous lesions and interstitial fibrosis on exposure to single- and multi-walled carbon nanotubes in rodents. They cite data indicating that inhaled long and thin multi-walled carbon nanotubes can move to the lining surrounding the lungs and penetrate it, where they can potentially cause mesothelioma. And they indicate that carbon nanotubes can act as a cancer promoter – with inhalation increasing the probability of developing lung cancer from exposure to other carcinogens.

While care is needed in extrapolating these results to real-world materials and exposures, Oberdörster and colleagues clearly indicate that, for some forms of the material (not all), at some exposure levels (as yet largely undetermined), there is a reasonable chance of adverse health effects occurring if exposure occurs. They also conclude that carbon nanotubes with similar physical and chemical characteristics will likely behave in similar ways; meaning that carbon nanotubes that look like MWNT-7, will probably behave like MWNT-7.

MWNT-7 occupies a gray zone between long, asbestos-like carbon nanotubes, and shorter nanotubes that do not behave like asbestos. However, Oberdörster *et al.* and others have indicated that, even when carbon nanotubes are not asbestos-like, they can elicit harmful responses. In particular, granuloma formation, fibrosis, and cancer promotion appear to be linked to the physical and chemical characteristics of inhaled nanotubes that are not necessarily long and straight. And there are indications that these responses are associated with both single nanotubes and aggregates of carbon nanotubes that can penetrate deep into the lungs – in other words, airborne particles that are respirable, or have an aerodynamic diameter less than approximately 5 μm.

And this brings us back to carbon nanotube spray-coatings such as Vantablack S-VIS.

Here, it is important to be clear that the carbon nanotubes in Vantablack S-VIS are not asbestos-like.  Rather, according to Surrey NanoSystems, they are short (less than 6 µm in length) and tightly bound together in a low-density matrix. Yet, they are used in a spray process that is likely to produce fine airborne droplets containing nanotubes. And they are applied to surfaces that could conceivably be touched, knocked, scratched, and potentially shed nanotube-containing particles.

Based on the current state of understanding, such applications do raise safety questions.  Yet, despite the potential of carbon nanotube materials to be harmful, everything hinges on the nature, form and concentrations of nanotube material that workers, users and others might actually be exposed to.  And here, the state of play is less clear.

What is known is that spray-paint processes can produce airborne particles smaller than 5 µm, and that without appropriate exposure controls, working with spray-on carbon nanotube paints could lead to respirable particles being inhaled – either during spraying, or from droplets suspended in the air after use.  It is also known that airborne carbon nanotube materials (single fibers and aggregates) are likely to be harmful at very low concentrations. The exposure limit currently recommended by NIOSH is 1 µg/m$^3$.[15] What is less well understood is whether spray-coated surfaces can shed particles containing nanotubes that are small enough to be inhaled (or ingested), and how safe or harmful these particles are likely to be.

Surrey NanoSystems indicates that, as far as they can tell, the chances of nanotube-containing particles being released from coated surfaces are low.  And they are careful to ensure that their product is not used where unacceptable exposures might occur. Responsible approaches like this will certainly help ensure products using spray-on carbon nanotubes remain safe, and will open the door to a world of innovative nanotechnology-based products.  Yet so much of this responsibility currently lies with manufacturers and users.

Considering this, I wonder what might happen when a less responsible company comes along with the next nanotube spray-paint.  Or when manufacturers, artists and others fail to understand how to use the paint appropriately.  Without doubt, carbon nanotubes can be used safely. But without greater awareness of the potential risks they present, it is by no means a foregone conclusion that they will be.